\documentclass[a4paper,fleqn]{cas-dc}

\usepackage[numbers]{natbib}
\usepackage{tikz}
\usetikzlibrary{shapes.geometric, arrows}

\tikzstyle{startstop} = [rectangle, rounded corners, minimum width=3cm, minimum height=1cm, text centered, draw=black, fill=red!30]
\tikzstyle{process} = [rectangle, minimum width=3cm, minimum height=1cm, text centered, draw=black, fill=orange!30]
\tikzstyle{arrow} = [thick,->,>=stealth]
\tikzstyle{annotation} = [text width=3cm, text centered, font=\small]

\usepackage{pgfplots}
\pgfplotsset{compat=1.18}
\def\tsc#1{\csdef{#1}{\textsc{\lowercase{#1}}\xspace}}
\tsc{WGM}
\tsc{QE}

\usepackage{algorithm}
\usepackage{algpseudocode}
\usepackage{tabularx}
\usepackage{float}

\begin{document}
\let\WriteBookmarks\relax
\def\floatpagepagefraction{1}
\def\textpagefraction{.001}

\shorttitle{Anomaly detection in network flows using unsupervised online machine learning}    

\shortauthors{}  

\title [mode = title]{Anomaly detection in network flows using unsupervised online machine learning}  



%

\author[1]{Alberto Miguel-Diez}[
    orcid=0009-0006-9969-3047
]

\cormark[1]


\ead{amigd@unileon.es}


\credit{Conceptualization, Methodology, Software, Validation, Investigation, Writing - Original Draft, Writing - Review \& Editing}


\author[1]{Adrián Campazas-Vega}[
    orcid=0000-0001-8237-5962
]


\ead{acamv@unileon.es}

\credit{Conceptualization, Methodology, Software, Validation, Investigation, Writing - Review \& Editing, Supervision}

\author[1]{{Á}ngel Manuel Guerrero-Higueras}[
    orcid=0000-0001-8277-0700
]

\ead{agueh@unileon.es}

\credit{Conceptualization, Methodology, Validation, Investigation, Writing - Review \& Editing, Supervision, Project administration, Funding acquisition}

\author[1]{Claudia {Á}lvarez-Aparicio}[
    orcid=0000-0002-7465-8054
]

\ead{calvaa@unileon.es}

\credit{Investigation, Writing - Review \& Editing}

\author[1]{Vicente Matellán-Olivera}[
    orcid=0000-0001-7844-9658
]

\ead{vmato@unileon.es}

\credit{Investigation, Writing - Review \& Editing, Supervision}

\affiliation[1]{organization={University of León},
            addressline={Robotics Group. Campus de Vegazana S/N.}, 
            city={León},
            postcode={24071}, 
            state={León},
            country={Spain}}

\cortext[1]{Corresponding author}



\begin{abstract}
Nowadays, the volume of network traffic continues to grow, along with the frequency and sophistication of attacks. This scenario highlights the need for solutions capable of continuously adapting, since network behavior is dynamic and changes over time. This work presents an anomaly detection model for network flows using unsupervised machine learning with online learning capabilities. This approach allows the system to dynamically learn the normal behavior of the network and detect deviations without requiring labeled data, which is particularly useful in real-world environments where traffic is constantly changing and labeled data is scarce. The model was implemented using the River library with a One-Class SVM and evaluated on the NF-UNSW-NB15 dataset and its extended version v2, which contain network flows labeled with different attack categories. The results show an accuracy above 98\%, a false positive rate below 3.1\%, and a recall of 100\% in the most advanced version of the dataset. In addition, the low processing time per flow (<0.033 ms) demonstrates the feasibility of the approach for real-time applications.
\end{abstract}


\begin{highlights}
\item An unsupervised online anomaly detection model is developed to dynamically learn normal network behavior from unlabeled netflow data.
\item The approach combines a One-Class SVM with a custom preprocessing and flow handling methodology, implemented using the River library.
\item It achieves 98.4\% accuracy, 100\% recall, and a false positive rate below 3.1\%, with average flow processing time under 0.033 ms.
\item Evaluations were conducted on NF-UNSW-NB15 and its extended version, covering 9 types of attacks.
\end{highlights}



\maketitle

\section{Introduction}\label{sec:introduction}

The use of technology has experienced steady growth in recent years. According to the We Are Social report, in January 2025 the number of people using the Internet reached 5.56 billion \cite{datareportal_digital_nodate}. Moreover, it is increasingly common for users to employ multiple devices simultaneously, such as mobile phones, tablets, or Internet of Things (IoT) devices. According to the Cisco Annual Internet Report, in North America the average number of devices and connections per capita is 13.4 \cite{cisco2020devices}.

As a result of this increased Internet usage, cybersecurity has taken on a fundamental role. The number of cyberattacks has risen considerably over time, reaching a historic record in the third quarter of 2024, when an average of 1,876 cyberattacks per organization was reported, representing a 75\% increase compared to the same period in 2023 and 15\% more than the previous quarter \cite{CheckPointBlog_2024}. This underscores the need for companies to implement security measures to protect their systems, such as firewalls, intrusion detection systems, or solutions based on artificial intelligence.


One of the main attack methods today are network-level attacks. These types of attacks typically leverage the Internet to carry out actions such as denial of service (DoS/DDoS), port scanning, vulnerability exploitation, or data exfiltration. In recent years, these threats have escalated significantly, both in volume and complexity. For example, DDoS attacks have become one of the most common forms of digital aggression, with year-on-year increases exceeding 90\% according to \cite{radware2024threat}. This percentage demonstrates that cyberattackers continue to employ tactics aimed at compromising network infrastructure, motivated by their effectiveness, scalability, and low cost compared to more sophisticated methods.

Historically, network traffic analysis was performed by examining the content of transmitted packets. A packet is the basic unit of data transmission in computer networks, including both the control information required for routing (headers) and the payload, i.e., the actual data that must reach the destination, such as part of a file, a message, or an HTTP request. However, this approach presents two major drawbacks: (i) it entails a considerable computational load, which becomes unfeasible in many modern technologies such as 5G due to the enormous number of packets transmitted in a very short time \cite{7387408}, and (ii) it relies on verifying whether packets match specific patterns to classify them as anomalous. This requires maintaining an updated database of signatures for all known attacks, which is difficult to sustain and does not allow for the detection of unknown threats such as zero-day attacks.

For this reason, flow-based analysis has gained increasing relevance. A network flow is defined as the set of packets exchanged between two devices during a communication within a given time interval. Instead of analyzing the content, the headers of these packets are examined to extract metrics such as the number of bytes sent, the number of packets, the protocol used, among others. This type of analysis makes it possible to identify malicious behaviors such as denial-of-service (DoS) attacks, botnets, or malware infections, as has already been demonstrated in the literature \cite{chen_exploring_2017,zeleke2025integratingexplainableaieffective}. Currently, there are different flow-level protocols, such as sFlow \cite{rfc3176} or IPFIX \cite{rfc3917}. However, one of the most widely used---since it is integrated by default in most commercial routers---is NetFlow \cite{rfc3954}, designed by Cisco.


Currently, the most effective solutions for detecting malicious traffic in network flows are based on machine learning techniques, both supervised and unsupervised. The supervised approach requires labeled datasets to train the models, which poses a challenge due to the high cost in terms of time and human effort, in addition to the difficulty of obtaining up-to-date data. Likewise, storing a large number of flows can become problematic in enterprise environments. In contrast, unsupervised models are trained with unlabeled data representing legitimate traffic and are capable of identifying anomalous behaviors without the need to know the type of attack in advance.


Delving deeper into unsupervised approaches, and considering that network traffic evolves constantly over time, a particularly relevant technique is online learning. This paradigm enables models to continuously adapt to new data without requiring prior training or the storage of large volumes of information. In enterprise networks, where traffic generation is constant and storing all flows can be technically expensive or even unfeasible due to privacy or performance constraints, this ability is especially advantageous. Moreover, since it does not require an initial training set, online learning allows for faster and more dynamic anomaly detection, improving responsiveness to emerging or previously unseen threats.

In this work, we develop an anomaly detection model based on unsupervised learning with online learning capabilities. The goal is for the model to continuously learn the behavior of the network where it is deployed and determine whether a given flow is anomalous, with the aim of detecting attacks.

To evaluate the performance of the model, we use the NF-UNSW-NB15 and NF-UNSW-NB15-v2 datasets, which consist of labeled NetFlow records and include nine different types of anomalies covering a wide range of attacks. Although the proposed approach relies on unsupervised learning and does not use labels during training, they are employed in the evaluation phase to verify whether the model can correctly identify anomalous behaviors. Thus, labels are used solely as a reference to measure effectiveness, without influencing the learning process itself.

The main contributions of this work are: (i) the design and implementation of an unsupervised model with online learning capabilities for detecting malicious traffic; (ii) the evaluation of the model using two versions of the NF-UNSW-NB15 dataset, which include multiple attack categories; and (iii) a methodology for processing flows in an online fashion.

The remainder of this paper is organized as follows. Section \ref{sec:related_work} summarizes the datasets commonly used in the literature to evaluate these models and the approaches proposed to date. Section \ref{sec:methodology} describes the methodology and tools employed. Section \ref{sec:evaluation} details the metrics used to evaluate the model and the hyperparameter optimization process. Section \ref{sec:results_discussion} presents and analyzes the results obtained. Finally, Section \ref{sec:conclusion} provides the conclusions of the study and outlines directions for future work.

\section{Related Work}
\label{sec:related_work}

As a preliminary step to the development of this research, a systematic literature review was conducted with the aim of identifying existing studies on anomaly detection in network flows using unsupervised techniques. Two main research questions were formulated: the first focused on determining the datasets employed for training and evaluating the algorithms, and the second on identifying the algorithms used for anomaly detection. The search process included all articles indexed in the IEEE Digital Library, Scopus, and Web of Science up to April 2024. The complete review can be consulted in \cite{miguel-diez_systematic_2025}.

In this review, 15 articles were analyzed in depth, revealing a total of 16 different datasets. Some were composed of real-world data, while the majority were generated synthetically in controlled environments. The five most frequently used datasets are summarized below:

The first is CIC-IDS-2017 \cite{sharafaldin_toward_2018}, created by the \textit{Canadian Institute for Cybersecurity}. It consists of PCAP files and also includes labeled network flows extracted with CICFlowMeter \cite{habibi_lashkari_characterization_2017}. The dataset covers a wide range of attacks, such as brute force (FTP and SSH), denial of service (DoS), Heartbleed, web attacks, infiltration, botnets, and distributed denial of service (DDoS). According to its authors, this dataset meets the eleven criteria proposed by Gharib et al. \cite{gharib_evaluation_2016} for the construction of a reliable benchmark dataset

The second dataset is the CTU-13 \cite{garcia_empirical_2014}. It was created by the CTU University in the Czech Republic in 2011. Its primary objective is to capture traffic associated with botnet activity, combined with both normal traffic and background traffic. The latter refers to flows whose nature is uncertain---whether they are benign or botnet-related. The dataset is available in both PCAP and NetFlow formats.

The MAWI dataset \cite{cho_traffic_2000}, developed by the \textit{MAWI Working Group}, provides real traffic traces from a Japanese academic network. It is characterized by containing unlabeled traffic, which poses a challenge for anomaly detection since clustering techniques or manual analysis are required to identify potential attacks. The traces are collected daily and made available in PCAP format, enabling their use with network flow processing tools.

The NSL-KDD dataset \cite{tavallaee_detailed_2009} is an improved version of the well-known KDD’99 dataset. It includes several attack categories such as DoS, R2L (Remote to Local), U2R (User to Root), and probing, allowing for the evaluation of intrusion detection algorithms across different scenarios.

Finally, the UNSW-NB15 dataset \cite{moustafa_unsw-nb15_2015} was developed by the \textit{Australian Centre for Cyber Security} (ACCS) with the aim of providing a more up-to-date and representative dataset of modern network threats. It was created in a realistic network environment, where both benign and malicious traffic were generated using the IXIA PerfectStorm tool.

This dataset includes nine attack categories: analysis, exploitation, backdoors, DoS, infiltration, injection, unauthorized access to systems, trojan attacks, and worms. It is available in both CSV and PCAP formats, making it suitable for models based on tabular features as well as for more detailed analyses of raw network traffic.



After analyzing the main datasets used in the literature, we now turn to a review of the most relevant unsupervised algorithms applied to anomaly detection in network flows. Over time, the proposed techniques have evolved from classical offline methods to hybrid approaches and online learning solutions.

Several recent studies have revisited traditional methods such as clustering and dimensionality reduction to address the problem of anomaly detection in network flows. For example, \cite{kabir_unsupervised_2020} applied the k-means algorithm to the KDD CUP’99 dataset \cite{uci_kdd}, achieving a detection rate of 0.998 and a false positive rate of 0.031. Subsequently, \cite{verkerken_unsupervised_2020} compared multiple approaches, including SVMs, Principal Component Analysis, and Isolation Forest. In this study, the Isolation Forest algorithm achieved the highest precision (0.9470), while the best recall value (0.9920) was obtained with the SVM classifier, highlighting the variability in performance depending on the metric employed.


In addition to these classical methods, techniques that combine statistical analysis with probabilistic models have also been explored. For instance, \cite{timcenko_hybrid_2022} introduced a hybrid approach that integrates entropy-based preprocessing with an Expectation-Maximization (EM) model, evaluated on the CTU-13 dataset \cite{GARCIA2014100}. However, the study does not provide quantitative metrics that would allow for a direct comparison of its performance with other proposals, which limits its objective assessment.


Another class of models frequently employed to address the problem discussed in this paper are Support Vector Machines (SVMs). In \cite{schueller_hierarchical_2018}, the authors proposed a hybrid hierarchical architecture that performs an initial analysis at the flow level and, if suspicious activity is detected, subsequently inspects the associated packets. This strategy achieved a detection rate of 83.81\% on the DARPA dataset \cite{lincoln_laboratory_1999_nodate}.


Some studies have focused on specific contexts such as the Internet of Things (IoT) or botnet detection. In \cite{sural_local_2022}, the Weighted Hamming Distance LID algorithm was employed to detect anomalies in IoT devices, achieving an AUROC value of 0.98 when applied to the Aposemat IoT-23 dataset \cite{garcia_iot-23_2020}. Along the same lines, \cite{chen_exploring_2017} proposed a botnet traffic detection system that combines Self-Organizing Maps, local outlier detection, and the KNN algorithm. The best performance was obtained with the latter, reaching a detection rate of 0.913 and a false positive rate of 0.051. On the other hand, BotMark \cite{wang_botmark_2020} introduced a hybrid approach based on network graphs and behavioral analysis for botnet detection, achieving an accuracy of 99.94\% using a proprietary dataset.





Despite the diversity of existing proposals, most of the approaches share a common characteristic: their offline nature. This limits their applicability in dynamic environments, where continuous model adaptation is required without the need for retraining. In this context, studies are beginning to emerge that adopt an unsupervised online learning perspective.


One example is ORUNADA \cite{dromard_online_2017}, a scalable online unsupervised method for anomaly detection in networks. The solution is based on a discrete time-sliding window to continuously update the feature space, combined with incremental grid clustering to rapidly detect anomalies. The system relies on network flow analysis, although it does not follow established standards such as NetFlow or sFlow. Finally, the evaluation was conducted using flows from a Spanish intermediate Internet service provider.



More recently, deep learning techniques have been explored in this context. In \cite{belacel_lstm_2022}, the authors proposed an encoder–decoder architecture based on recurrent neural networks with long short-term memory. Although the article focuses on anomaly detection in a general sense, one of the datasets used for evaluation was KDD CUP’99, on which the model achieved an AUC of 0.97 for anomalous HTTP traffic detection, but significantly lower performance (AUC of 0.53) for SMTP traffic. Finally, the work of \cite{odiathevar_online_2022} introduced a hybrid framework: an offline deep learning model is used to extract features of normal (benign) traffic, providing a bias for an online outlier detection model to select data for training. The authors evaluated their framework using three deep learning models (AE, VAE, DAE) and four outlier detection methods (IOCSVM, IGMM, OCSVM, GMM, LOF, KDE) on the UNSW-NB15 and CTU-13 datasets. The best results with their hybrid approach, when evaluated on the UNSW-NB15 dataset, were obtained with the AE+MW+IOCSVM model, achieving an AUC of 0.9506.





Despite the large number of proposals in the literature for anomaly detection in network flows, most rely on offline approaches that require all data to be available beforehand and cannot adapt to traffic changes in real time. Conversely, the few proposals that apply unsupervised online learning present significant limitations: either they do not follow standards widely adopted in real-world environments (such as NetFlow), or they lack comprehensive validation on datasets with multiple types of attacks. In this context, the solution proposed in this work represents a novel contribution by combining three key features that are rarely integrated simultaneously: unsupervised learning, online processing, and standardized flow analysis. This combination enables effective anomaly detection without the need for labeled data or periodic retraining, while maintaining a computational cost low enough to be deployed in real-time systems.

\section{Methodology}\label{sec:methodology}

This section presents the methodology followed to design, implement, and evaluate an anomaly detection model for network flows using unsupervised machine learning techniques with online learning capabilities. It describes the foundations of the online learning paradigm, the datasets employed for evaluation, the preprocessing steps required to adapt flows to the model, and finally, the algorithms used along with their configuration.

\subsection{Online Machine Learning}

Before delving into the details of online learning, it is important to discuss traditional machine learning algorithms. Batch (or offline) learning is divided into two phases: training and validation. To carry out this process, the entire dataset must be available beforehand, which in some cases can be problematic in systems with limited storage capacity. In addition, training typically requires high computational resources and does not allow the model to easily adapt to real-time changes. This type of approach also entails several additional drawbacks \cite{bartz_introduction_2024}:

\begin{itemize}
    \item \textbf{Memory requirements}: this issue arises when the size of the dataset exceeds the amount that can be stored in RAM.
    \item \textbf{Concept drift}: this occurs when models become outdated because the relationships established by the algorithm are no longer valid. This leads to performance degradation and, therefore, requires retraining the model. For example, if the typical traffic in a corporate network changes due to the adoption of new cloud applications or internal restructuring, the usual traffic patterns will change, rendering a previously trained model ineffective.
    \item \textbf{Unknown data}: the model cannot easily learn from new data that contains previously unseen attributes. When features appear that were not present in the original training set, the model cannot directly integrate them into its existing knowledge. This forces retraining the model from scratch with a dataset that combines both historical and new data. In production environments, such as anomaly detection systems, this process is performed in isolation, and once retraining is complete, the new model replaces the old one.
\end{itemize}

In contrast, Online Learning is a strategy in which the model learns continuously as data streams are received \cite{shahraki_comparative_2022}. As a result, each training step is faster and requires lower computational cost. This brings two main advantages: (i) devices with limited memory can implement these models \cite{gepperth_incremental_2016}, and (ii) the model can quickly adapt to new data as it arrives.


As with traditional machine learning, online learning can also be performed in supervised, unsupervised, and semi-supervised modes \cite{shahraki_comparative_2022}. This depends on whether a labeled dataset is available beforehand. In this work, we opted for an unsupervised approach. This choice is motivated by the fact that obtaining a labeled dataset with various types of anomalies is difficult, since in real network environments the collection of labeled data requires significant manual effort. Furthermore, threats evolve continuously, which can quickly render a labeled dataset obsolete.

In unsupervised online learning, data \( D = (x_1, x_2, \dots, x_m) \) arrive sequentially without associated labels. The goal is to construct a model \( F \approx p(y \mid X) \) that can identify patterns or anomalies as new instances \( x_t \) are received.

Unlike batch learning, where data are available in advance, online learning updates the model incrementally without the need to store large volumes of data. At each step \( t \), the model \( F_t \) is built using the new instance \( x_t \) together with the previous state \( F_{t-1} \).

In addition, within the unsupervised learning paradigm, a technique known as \textit{novelty detection} is employed. This approach trains the model exclusively on examples representing normal system behavior, with the aim of identifying patterns that deviate significantly from the learned behavior \cite{PIMENTEL2014215}. In other words, the model does not attempt to classify different types of anomalies but rather detects when a new observation differs from what has been seen during training. This strategy is particularly useful in environments where anomalies are rare, diverse, or difficult to label, as is often the case in real network traffic.

Although online learning models do not require prior training, a short warm-up phase is typically performed before the first predictions in order to achieve better performance and to allow comparisons with batch machine learning models \cite{bartz_experimental_2024}.

Another fundamental aspect of any machine learning pipeline is feature normalization or scaling, as it often improves performance \cite{DEAMORIM2023109924}. This process consists of transforming input variables so that they are comparable in scale, typically with zero mean and unit variance, or within a bounded range such as [0, 1]. The need for this preprocessing arises because many machine learning algorithms, such as Support Vector Machines, are sensitive to the magnitude of variables \cite{pinheiro2025impactfeaturescalingmachine}. If some features have much larger numerical values than others, they may dominate the model’s behavior and introduce unwanted biases into the learning process.

In the case of online learning, scaling must also be performed incrementally, since the entire dataset is not available in advance. This implies that the scaler’s parameters---such as mean, variance, or extreme values---must be dynamically updated as new data arrive. Unlike batch learning, where a global scaling can be precomputed over the full dataset, in online learning the model must adapt to changes in the data distribution without reprocessing past observations.

Nevertheless, to prevent early instances from distorting the scaler’s parameters, a brief warm-up phase is required, using a small subset of flows representative of the network’s normal traffic. This step properly initializes the scaling model before the anomaly detector warm-up, ensuring greater stability and consistency in the preprocessing stage.

To implement all these processes, this work employs the \texttt{River} library~\cite{montiel_river_2021}. This library provides a collection of algorithms tailored to the online learning paradigm, both for predictive models and preprocessing tasks.

\subsection{Dataset}


Unsupervised learning algorithms do not require labeled data for training; however, labels are necessary when evaluating the effectiveness of the proposed models. According to the previously conducted literature review \cite{miguel-diez_systematic_2025}, the UNSW-NB15 dataset \cite{moustafa_unsw-nb15_2015} is among the most widely used for assessing anomaly detection systems. In addition, it includes a large number of diverse attack types, making it particularly suitable for evaluating models of this kind. The dataset consists of 2,218,761 benign flows and 321,283 attack flows.

For this work, the NF-UNSW-NB15 dataset \cite{deze_netflow_2021} was used, which is a flow-based version of the well-known UNSW-NB15 dataset. While the original dataset consists of full packet captures in PCAP (Packet Capture) format, NF-UNSW-NB15 transforms this information into flow records following the NetFlow version 9 standard \cite{benoit_claise_cisco_2004}. In this case, 12 features were selected, capturing both generic information and protocol-specific details (such as DNS or FTP). Table~\ref{tab:NF-UNSW-NB15-fields} lists the selected features and their descriptions. Additionally, the dataset is complemented with two extra attributes: a label indicating whether a flow is benign or anomalous, and the attack category in the case of malicious traffic.

\begin{table}[ht]
\centering
\caption{List of NetFlow version 9 fields used in NF-UNSW-NB15 \cite{deze_netflow_2021}}
\begin{tabular}{p{3.6cm}|l} 
\toprule
\textbf{Feature} & \textbf{Description} \\
\midrule
IPV4\_SRC\_ADDR & IPv4 source address \\
IPV4\_DST\_ADDR & IPv4 destination address \\
L4\_SRC\_PORT & IPv4 source port number \\
L4\_DST\_PORT & IPv4 destination port number \\
PROTOCOL & IP protocol identifier byte \\
TCP\_FLAGS & Cumulative of all TCP flags \\
L7\_PROTO & Layer 7 protocol (numeric) \\
IN\_BYTES & Incoming number of bytes \\
OUT\_BYTES & Outgoing number of bytes \\
IN\_PKTS & Incoming number of packets \\
OUT\_PKTS & Outgoing number of packets \\
FLOW\_DURATION\_\\MILLISECONDS & Flow duration in milliseconds \\
\bottomrule
\end{tabular}
\label{tab:NF-UNSW-NB15-fields}
\end{table}

In addition, the model was also evaluated using an extended version of the original dataset, referred to as NF-UNSW-NB15-v2 \cite{sarhan_towards_2022}. Although the same features as in the previous version were employed in this work to ensure consistency in the experiments, this variant introduces differences in preprocessing and in the structure of the flows. Its inclusion therefore makes it possible to analyze whether the model’s performance remains consistent under slight variations in the data source, providing a more comprehensive view of its applicability across different network scenarios.

\begin{table}
\caption{Distribution of flows in the NF-UNSW-NB15 and NF-UNSW-NB15-v2 datasets}\label{distribution-dataset}
\begin{tabular*}{\tblwidth}{@{}LLL@{}}
\toprule
\textbf{Class} & \textbf{NF-UNSW-NB15} & \textbf{NF-UNSW-NB15-v2}\\
\midrule
Benign & 1,550,712   & 2,295,222   \\
Fuzzers & 19,463  & 22,310  \\
Analysis & 1,995  & 2,299  \\
Backdoor & 1,782  & 2,169  \\
DoS & 5,051  & 5,794  \\
Exploits & 24,736   & 31,551  \\
Generic & 5,570   & 16,560  \\
Reconnaissance & 12,291   & 12,779  \\
Shellcode & 1,365  & 1,427  \\
Worms & 153   & 164  \\
\bottomrule
\end{tabular*}
\end{table}

The dataset is imbalanced, meaning that the number of flows considered ``normal'' is larger than the number of malicious flows. In NF-UNSW-NB15, the total number of flows is 1,623,712, of which 72,406 (4.46\%) correspond to attacks. In the v2 version, the total number of flows increased significantly, reaching 2,390,275, with an attack proportion of 3.98\%. Table \ref{distribution-dataset} shows the number of flows in each class of the dataset.


\subsection{Preprocessing Pipeline}
\label{subsec:preprocessing_pipeline}

\begin{figure*}
    \centering
    \includegraphics[width=0.8\linewidth]{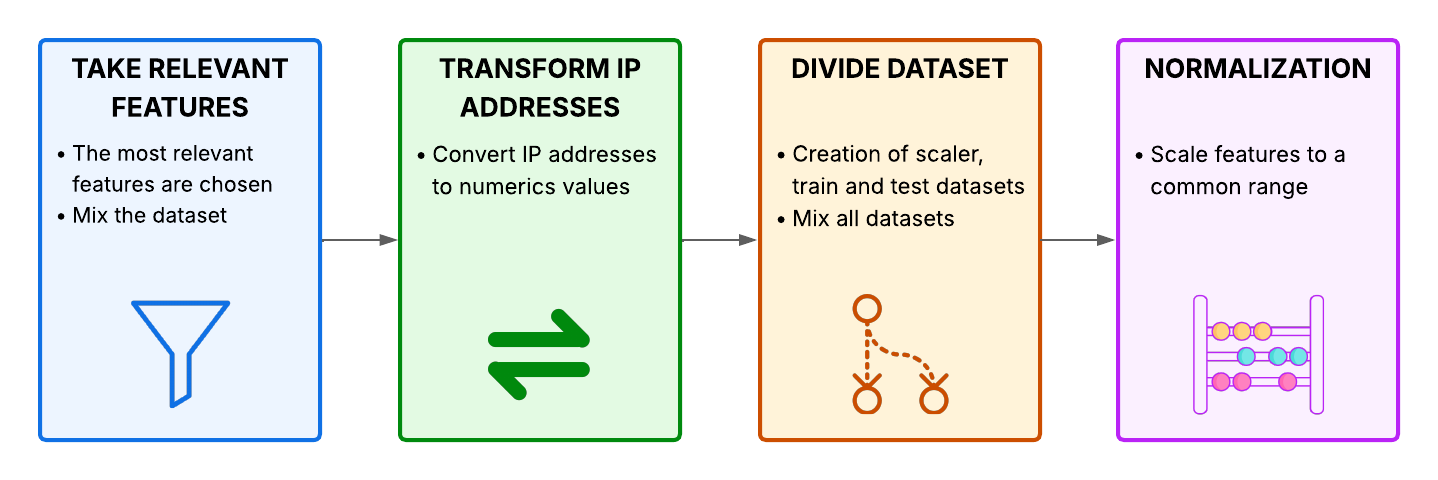}
    \caption{Four-stage preprocessing pipeline applied to network flows before feeding the model.}
    \label{fig:preprocessing-image}
\end{figure*}

Before feeding the flows into the model, a preprocessing phase is carried out. This process is divided into four steps, as shown in Figure \ref{fig:preprocessing-image}. Each of these stages is described in detail below.

 The first phase consists of selecting the most relevant features. In this case, these are the source and destination IP addresses, ports, the protocol used, incoming and outgoing bytes, and finally, the flow duration in milliseconds. These features were chosen because, after several analyses, they were found to provide the best results. In addition, the dataset is then shuffled randomly to verify that the online model produces consistent results regardless of the order in which the flows are received.

In the second phase, IP addresses---which in the datasets found in the literature are IPv4---are converted into integers. This step is performed using Python’s \verb|ipaddress| library. The conversion is based on interpreting the four octets of the address as coefficients in base 256, so that an address of the form \texttt{a.b.c.d} is transformed according to the following expression:

\begin{equation}
    \texttt{IP}_\texttt{integer} = a \times 256^3 + b \times 256^2 + c \times 256^1 + d \times 256^0
\end{equation}

The reason for this transformation is that the model to be used only works with integers; therefore, any features that do not meet this condition must be modified.

Subsequently, in the third phase, the dataset is divided into three subsets: scaler, training, and test. The scaler dataset consists of 1,000 flows, whose purpose is to train the scaler before feeding the training flows into the model. The number of flows used for scaling was hyperparameterized, with this value providing the best results. The training dataset consists of 100,000 benign flows. The aim is to use the smallest possible amount before evaluating the model, and this number was found to yield satisfactory results.


The test set is composed of all anomalous flows present in the dataset, together with an equivalent number of benign flows. Although in a real network environment the proportion of anomalous traffic is significantly lower than benign traffic, a balanced evaluation set was created in order to avoid bias in the performance metrics. This is particularly important for metrics such as accuracy, which can be affected by a disproportionate class distribution, ensuring that the results more reliably reflect the model’s ability to distinguish between normal and anomalous traffic. Finally, both subsets are randomly shuffled to guarantee that the order in which flows arrive does not influence the behavior of the online model. Table~\ref{tab:flow-distribution} shows the number of flows used in each phase of the experimental process after applying balancing to the evaluation set.

\begin{table}[h]
\caption{Number of flows used in each phase}
\label{tab:flow-distribution}
\centering
\begin{tabular}{lccc}
\toprule
\textbf{Dataset} & \textbf{Scaler} & \textbf{Training} & \textbf{Evaluation} \\
\midrule
NF-UNSW-NB15 & 1,000 & 100,000 & 144,812 \\
NF-UNSW-NB15-v2 & 1,000 & 100,000 & 190,106 \\
\bottomrule
\end{tabular}
\end{table}

In the fourth phase of the process, all features are normalized to ensure homogeneity in their magnitude. For this purpose, the following scaling algorithms from the River library \cite{montiel_river_2021} were considered: MinMaxScaler, MaxAbsScaler, Normalizer, RobustScaler, and StandardScaler.

Online scaling algorithms adjust their parameters incrementally as new data arrive, allowing continuous adaptation without the need to store large volumes of information. MinMaxScaler transforms values into the \([0,1]\) range by updating the observed minimum and maximum. MaxAbsScaler rescales according to the maximum absolute value of each feature, preserving the structure of the data with values centered around zero. Normalizer adjusts each sample individually to unit norm, which is useful in scenarios where the relative scale within each sample is relevant. RobustScaler employs robust statistics such as the median and interquartile range, making it less sensitive to outliers. Finally, StandardScaler applies online normalization based on mean and standard deviation, ensuring an approximately normalized distribution without requiring prior knowledge of the data.


Algorithm \ref{alg:preprocess} summarizes the entire preprocessing procedure, which consists of creating the different training and validation datasets, normalizing the features, and transforming data types that are not compatible with the model.

\begin{algorithm}
\caption{Flow Preprocessing}
\label{alg:preprocess}
\begin{algorithmic}[1]
\State \textbf{Input:} Network flow dataset $X$

\State $X\gets$ Shuffle the dataset

\State $X\gets$ Select relevant columns from $X$

\State $X\gets$ Transform IP addresses in $X$ to integer format

\State $X_{scaler}\gets$ Take 1,000 benign flows from $X$

\State $X_{train}\gets$ Take 100,000 benign flows from $X \setminus X_{scaler}$

\State $X_{test}\gets$ Balance benign and anomalous flows from $X \setminus (X_{scaler} \cup X_{train})$

\State \textbf{Scaler Training Phase:}
\State Initialize the scaler $S$

\State $X_{scaler}\gets$ Shuffle $X_{scaler}$

\For{each flow $f$ in $X_{\text{scaler}}$}
    \State Train scaler $S$ with $f$
\EndFor

\State \textbf{Feature Normalization Phase:}
\State $X_{train}\gets$ Shuffle $X_{train}$
\State $X_{test}\gets$ Shuffle $X_{test}$

\For{each flow $f$ in $X_{train} \cup X_{test}$}
    \State Update scaler $S$ with $f$
    \State Transform flow $f$ using $S$
\EndFor
\end{algorithmic}
\end{algorithm}

\subsection{Algorithms}

For anomaly detection in network flows, the One-Class SVM (OCSVM) model was employed in its online learning variant provided by the River library \cite{montiel_river_2021}. This model is widely used in novelty detection tasks, as it can learn the distribution of normal data and detect significant deviations as potential anomalies. Previous studies using the model in offline settings have reported good results \cite{Campazas-Vega2023-ec,anomalyGeorge2012}; therefore, it is of particular interest to evaluate its performance when applied in an online context.

Unlike the standard OCSVM in batch learning, the online implementation in River updates the model incrementally with each new flow received, without the need to store large volumes of data. This is essential in network anomaly detection environments, where data are generated in real time and the model must continuously adapt to potential changes in legitimate traffic.

The model is initially trained with a set of benign flows, allowing it to learn the normal structure of network traffic. Subsequently, as new data arrive, it incrementally adjusts its separating hyperplane according to the new observations.

Several key parameters were tuned to optimize performance. In particular, the parameter \(\nu\) controls the estimated proportion of anomalies in the data. In addition, \verb|InverseScaling| was employed as the learning rate scheduler to update the bias term (intercept). This method gradually reduces the learning rate in a power-based manner, ensuring a progressive adaptation of the model. If an initial learning rate \(\eta\) is defined, the learning rate at step \(t\) is expressed as:

\begin{equation}
    \eta_{t+1} = \frac{\eta}{(t+1)^p}
\end{equation}

where \(p\) is a user-defined configurable parameter, for which the default value provided by the River library was used.

It is worth noting that this type of model does not provide a direct binary classification; instead, it outputs a continuous score, where higher values indicate a greater likelihood that a flow is anomalous. However, this score is not bounded within a specific interval, which makes it difficult to define a fixed decision threshold. To address this limitation, a quantile-based filter was employed, which classifies a sample as benign or anomalous according to its relative position within the observed score distribution. In this way, a dynamic threshold is defined that automatically adapts to the data, enabling more robust detection in the presence of variations in network traffic. This filter takes as a parameter $q$, which represents the quantile level above which a score is classified as anomalous, and this parameter was tuned during hyperparameter optimization.

Next, Algorithm \ref{alg:detec_nov} presents the pseudocode used in this work to apply an online learning model for anomaly detection. In this case, although training is not strictly required for this type of model, a warm-up phase with 100,000 flows was performed to allow the model to converge properly.

\begin{algorithm}
\caption{Implementation of a novelty detection model}
\label{alg:detec_nov}
\begin{algorithmic}[1]
\State \textbf{Input:} Preprocessed dataset $X$, novelty detection model $M$

\State Initialize model $M$

\State $X_{\text{train}}\gets$ Obtain a set of $n$ benign samples from $X$
\State $X_{\text{test}}\gets$ $X \setminus X_{train}$

\State \textbf{Training Phase (\textit{Warm-up}):}
\For{each data point $x$ in $X_{\text{train}}$}
    \State Train model $M$ with $x$
\EndFor

\State \textbf{Training–Validation Phase:}
\For{each data point $x$ in $X_{\text{test}}$}
    \State $\text{score}\gets$ Obtain anomaly prediction for $x_{\text{test}}$
    \State $\text{anomaly}\gets$ Classify the flow based on the score
    \If {not anomaly}
        \State Train model $M$ with $x$
    \EndIf
\EndFor

\end{algorithmic}
\end{algorithm}

\section{Evaluation Setup}\label{sec:evaluation}

Once the anomaly detection model has been designed and implemented, it is necessary to establish a rigorous procedure to evaluate its performance. This section outlines the metrics used, and also details the hyperparameter optimization process as well as the hardware employed in the experiment.

\subsection{Metrics}
\label{subsec:metrics}

The problem addressed in this paper is a binary classification task (``benign'' or ``anomalous''). To evaluate the performance of the model, we employ metrics derived from the confusion matrix. This matrix is built from four key values: True Positives (TP), True Negatives (TN), False Positives (FP), and False Negatives (FN). True Positives (TP) represent the cases in which the model correctly predicts that an instance is ``anomalous'' and it actually is. True Negatives (TN) are those cases where the model correctly predicts that an instance is ``benign'' and it indeed is. False Positives (FP), on the other hand, correspond to cases where the model incorrectly classifies an instance as ``anomalous'' when it is actually ``benign''. Finally, False Negatives (FN) are cases where the model incorrectly classifies an instance as ``benign'', when in fact it should have been classified as ``anomalous''.

The following metrics are used to evaluate the performance of the model:

\textit{Accuracy} indicates the frequency with which the model makes correct predictions with respect to the total number of predictions.

\begin{equation}
    Accuracy = \frac{TP +  TN}{TP + TN + FP + TN}
\end{equation}

\textit{Precision} is a metric that indicates the proportion of correct positive predictions with respect to the total number of positive predictions made by the model. In other words, it measures how accurate the model’s positive predictions are.

\begin{equation}
    Precision = \frac{TP}{TP + FP}
\end{equation}

\textit{Recall} is a metric that measures the proportion of true positives correctly identified by the model with respect to the total number of actual positive instances. In other words, recall indicates the model’s ability to find all positive examples.

\begin{equation}
    Recall = \frac{TP}{TP + FN}
\end{equation}

The \textit{false positive rate} (FPR) is a metric that indicates the proportion of instances that have been incorrectly classified as positive out of the total number of instances that are actually negative.

\begin{equation}
    FPR = \frac{FP}{FP + TN}
\end{equation}

Finally, the \textit{F1-Score} is a metric that combines precision and recall into a single measure, providing a balanced evaluation of the model’s performance, particularly when there is an imbalanced distribution between positive and negative classes. The F1-Score is defined as the harmonic mean of precision and recall, which means that it accounts for both false positives and false negatives.

\begin{equation}
    f_1score = 2 \times \frac{precision \times recall}{precision + recall}
\end{equation}

\subsection{Hyper-parameter Optimization}
\label{subsec:hyper-parameter-optimization}

According to its official documentation, the River library does not include specific functions for model hyperparameter tuning, unlike more common libraries such as scikit-learn \cite{pedregosa_scikit-learn_2011}, which provide utilities such as \verb|GridSearchCV| or \verb|RandomizedSearchCV| to optimize a model with the aim of maximizing its performance based on specific metrics, such as those described in Section \ref{subsec:metrics}.

In this work, a custom script was developed to perform hyperparameter optimization of the model, following the workflow detailed in Section \ref{subsec:preprocessing_pipeline}. The script allows adjusting the values of the model parameters \(\nu\) and \(\eta\). It also enables testing different scaling algorithms, such as MaxAbsScaler, MinMaxScaler, Normalizer, RobustScaler, and StandardScaler. In addition, it provides an easy way to configure the number of flows used in the warm-up phase, the features employed in the analysis, and the handling of IP addresses.

Since online learning models update their parameters sequentially as new data arrive, the order in which flows are presented can directly affect the results obtained. For this reason, an additional script was created to execute the hyperparameter optimization multiple times ($n$  runs), randomly varying the order of the flows in each iteration. This makes it possible to obtain more robust estimates of the model’s performance and evaluate its stability under different input sequences.

Furthermore, plots are generated to visualize the evolution of the main metrics, such as accuracy, precision, and recall, with the aim of ensuring that the model’s performance does not degrade over time. The source code of this work is publicly available in a GitHub repository \cite{Miguel-Diez_Anomaly_detection_in_2025}.

\subsection{Hardware Setup}


The experiments were conducted on a machine equipped with an AMD Ryzen 9 7900X processor with 12 cores and 32 GB of RAM. The operating system used was Ubuntu 24.04.1 LTS (64-bit), running the Linux kernel version 6.8.0-52-generic. No graphics processing unit (GPU) was used for the experiments.

\section{Results and Discussion}
\label{sec:results_discussion}

This section presents and analyzes the results obtained from the proposed anomaly detection model. It is structured as follows: first, the global performance metrics achieved after the hyperparameter optimization process are presented, offering an overview of the model’s behavior across all scenarios. Subsequently, a detailed discussion is provided for each dataset used in the evaluation, highlighting specific behaviors, model performance variations, and potential explanations for the observed differences.

The results are summarized in Table~\ref{tab:resultados-modelos}, which presents the performance metrics obtained by the model after the hyperparameter optimization process. These values correspond to the average of 12 independent executions, carried out using the custom optimization script described in Section~\ref{subsec:hyper-parameter-optimization} and following the methodology detailed in Section~\ref{sec:methodology}.

\begin{table*}
\caption{Results obtained with the UNSW-NB15 dataset family}
\begin{tabular*}{\tblwidth}{@{}LLLLLLLL@{}}
\toprule
Dataset & Accuracy & Precision & Recall & FPR & F1-Score & Training (s) & Inference (s)\\
\midrule
NF-UNSW-NB15 & 0.9853 & 0.9745 & 0.9967 & 0.0261 & 0.9854 & 1.0688 & 4.7179 \\
NF-UNSW-NB15-v2 & 0.9848 & 0.9706 & 1.0 & 0.0304 & 0.9850 & 1.0177 & 5.4276 \\

\bottomrule
\end{tabular*}
\label{tab:resultados-modelos}
\end{table*}

The accuracy obtained with both datasets is remarkably high---around 98\%---demonstrating the effectiveness of online learning methods for anomaly detection. Moreover, the model achieved a precision close to 97\%, indicating its ability to correctly identify the vast majority of anomalous flows. This suggests that the model is capable of distinguishing between the different types of attacks described in Table~\ref{distribution-dataset}. In particular, for NF-UNSW-NB15-v2, the model reaches a perfect recall of 1.0, successfully identifying all anomalies present in the dataset.

In contrast, the false positive rate (FPR) remains low in both cases---2.61\% and 3.04\% respectively---which is a crucial factor for real-world applications. In industrial environments, a high FPR would result in an excessive number of false alarms, decreasing operational efficiency and potentially overwhelming security analysts. The low FPR reported here makes the deployment of the model in production environments feasible.

Another important aspect of the evaluation is processing time, especially when considering real-time intrusion detection. In the case of NF-UNSW-NB15, the recorded execution times are as follows:

\begin{itemize}
    \item Initial training time (100{,}000 warm-up flows): \( 1.0688 \) seconds.
    \item Total evaluation time (including prediction and online updates): \( 4.7179 \) seconds.
    \item Average processing time per flow: \( 0.0326 \) ms.
\end{itemize}

These values demonstrate that the model is efficient enough to be deployed in real-time scenarios, maintaining a balanced trade-off between predictive accuracy and computational cost.

Similarly, for the NF-UNSW-NB15-v2 dataset, the model yielded the following performance:

\begin{itemize}
    \item Initial training time (100{,}000 warm-up flows): \( 1.0177 \) seconds.
    \item Total evaluation time (including prediction and online updates): \( 5.4276 \) seconds.
    \item Average processing time per flow: \( 0.0285 \) ms.
\end{itemize}

A slight increase in total evaluation time is observed compared to the first dataset. However, this is expected due to the higher number of flows processed, 45{,}294 more in the evaluation phase. Despite this increase, the average processing time per flow remains below 0.033 ms, well within acceptable limits for real-time detection.

\begin{figure*}[H]
    \centering
    \includegraphics[width=0.98\linewidth]{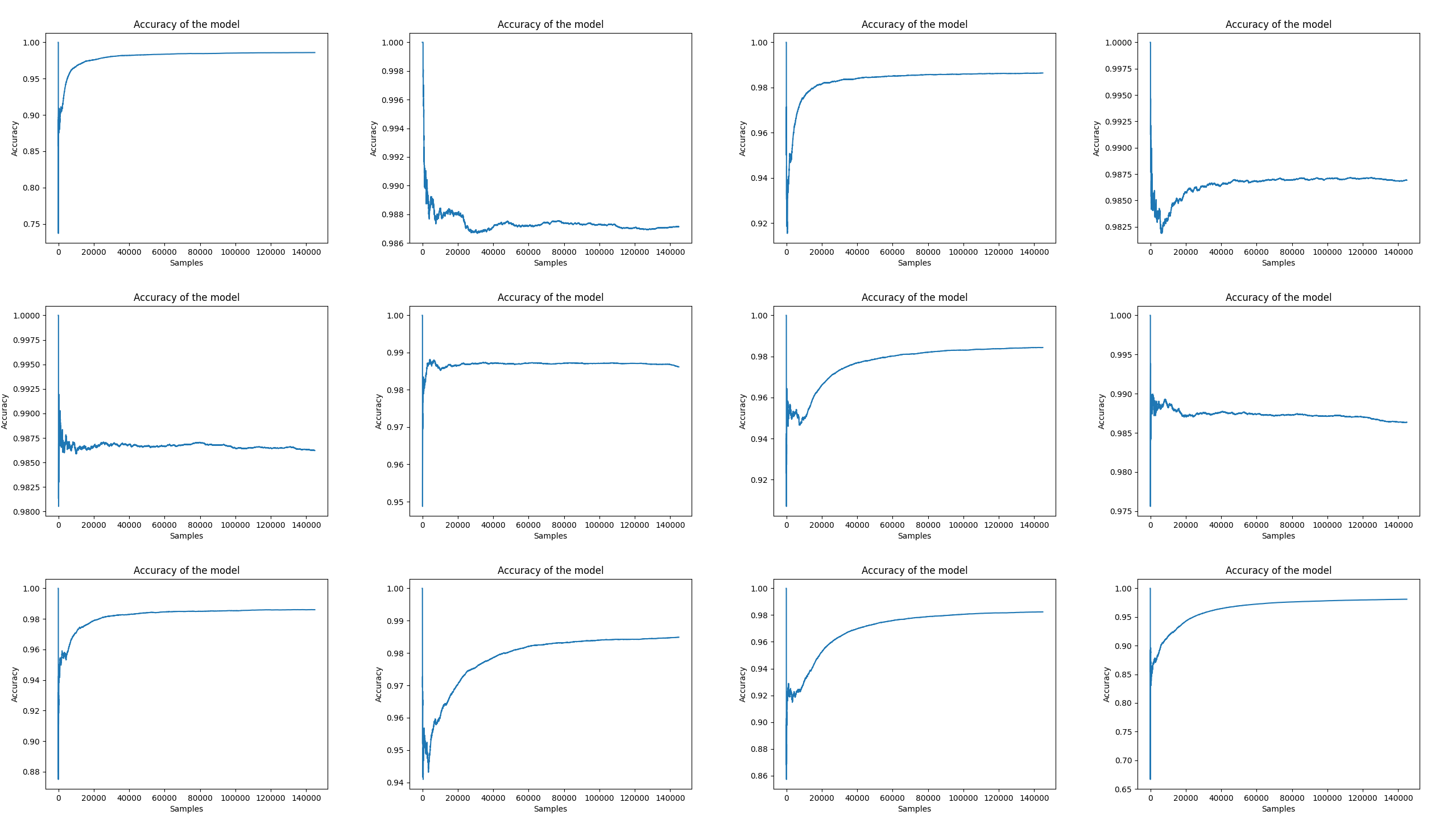}
    \caption{Evolution of accuracy throughout the evaluation phase on the NF-UNSW-NB15 dataset (12 runs)}
    \label{fig:accuracy-nb15-v1}
\end{figure*}

\begin{figure*}[htb]
    \centering
    \includegraphics[width=0.98\linewidth]{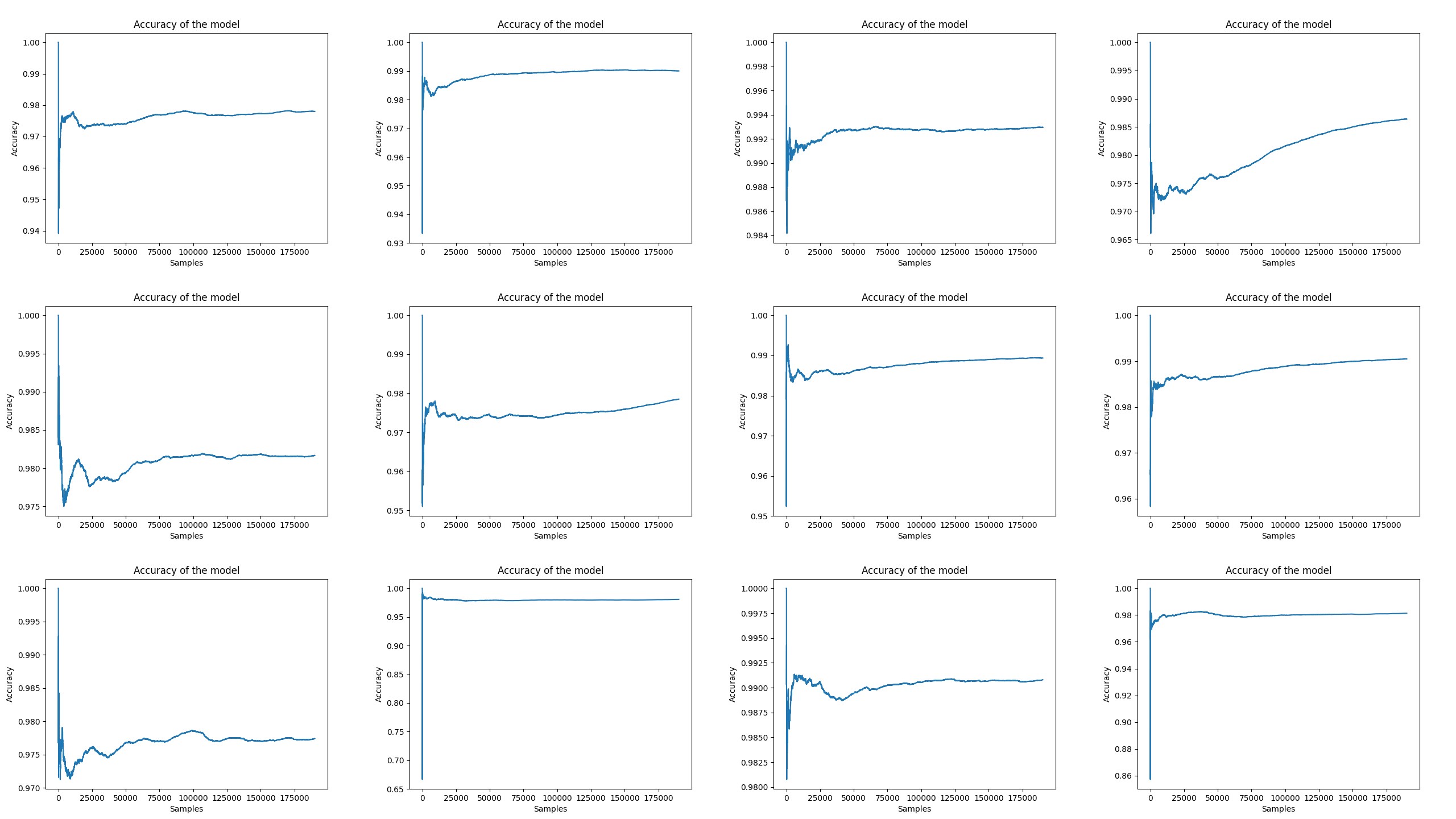}
    \caption{Evolution of accuracy throughout the evaluation phase on the NF-UNSW-NB15-v2 dataset (12 runs)}
    \label{fig:accuracy-nb15-v2}
\end{figure*}

Furthermore, the evolution of the \textit{accuracy} metric throughout the evaluation phase exhibits consistent behavior in both datasets, with an initial adjustment period followed by performance stabilization (see Figures~\ref{fig:accuracy-nb15-v1} and~\ref{fig:accuracy-nb15-v2}). This reflects the model’s adaptability in an online learning context, where it incrementally adjusts its decision boundary as new data arrives. This behavior is particularly valuable in dynamic environments, where network traffic patterns evolve over time.

It is also worth noting that, unlike traditional machine learning models that freeze their parameters during evaluation, the proposed model continues learning in real time. This allows it to adapt to new benign patterns and potentially unseen threats, enhancing robustness. Even with this additional computational step, the system maintains minimal latency, reinforcing its viability for operational deployment.

The following paragraphs present the specific hyperparameters that yielded the best results for each dataset. These configurations complement the previous analysis by providing insight into the internal setup that enabled the model to achieve such performance levels.

Table~\ref{tab:params-nfunsw} shows the hyperparameter configuration that yielded the best results for the NF-UNSW-NB15 dataset. The selected values reflect a balanced setup that optimizes detection accuracy while maintaining a low false positive rate. Specifically, the quantile threshold \( q = 0.99 \) defines a high-sensitivity decision boundary, meaning only the top 1\% of flows with the most extreme anomaly scores are flagged. This conservative threshold is typical in novelty detection settings, where minimizing false alarms is essential. The parameter \( \nu = 0.05 \) estimates the proportion of expected anomalies in the data; lowering this value would make the model stricter, potentially missing subtle anomalies, while increasing it could lead to excessive false positives. Finally, the learning rate \( \eta = 0.1 \) controls how quickly the model adapts to new data. A moderate value ensures that the model can adjust to evolving traffic without overreacting to noise. Together with the use of the \textit{MaxAbsScaler}, which is robust to large-scale variations and preserves sparsity, this configuration enables stable and efficient online learning.

The \textit{MaxAbsScaler} was selected as the normalization method due to its suitability for streaming data scenarios such as network flow analysis. This scaler transforms each feature by dividing its value by the maximum absolute value observed so far, effectively mapping the data into the range \([-1, 1]\) without shifting the mean. In the online setting, this process is defined incrementally as follows:

\begin{equation}
    x^{(t)}_{\text{scaled}} = \frac{x^{(t)}}{M^{(t)}}, \quad \text{with} \quad M^{(t)} = \max(M^{(t-1)}, \lvert x^{(t)} \rvert)
    \label{eq:maxabs_online}
\end{equation}

Here, \( x^{(t)} \) represents the input value at time step \( t \), and \( M^{(t)} \) denotes the maximum absolute value observed up to that point. This recursive formulation, shown in Equation~\ref{eq:maxabs_online}, allows the scaler to adapt continuously to new data without the need to store or access the full dataset history.

Unlike methods such as \textit{StandardScaler}, it does not center the data, which helps preserve sparsity~\cite{jakupov2023phd}. In the context of network flows, where features often vary widely in scale and include bursty traffic patterns, \textit{MaxAbsScaler} ensures that all features contribute proportionally without distorting the relative magnitude between them.

This variety of scaler has previously been employed in other studies, like \cite{Haclar2024,SAMANTARAY2024100478,Okey2022}, to identify anomalies in network flows. Consequently, it's unsurprising that it performs effectively with online learning as well.

\begin{table}[h]
\caption{Model configuration for NF-UNSW-NB15 dataset}
\label{tab:params-nfunsw}
\centering
\begin{tabular}{ll}
\toprule
\textbf{Parameter} & \textbf{Value} \\
\midrule
Scaler & MaxAbsScaler \\
Quantile threshold \( q \) & 0.99 \\
Anomaly rate parameter \( \nu \) & 0.05 \\
Learning rate \( \eta \) & 0.1 \\
\bottomrule
\end{tabular}
\end{table}

Similarly, the model was evaluated using the NF-UNSW-NB15-v2 dataset with the hyperparameter configuration presented in Table~\ref{tab:params-nfunsw-v2}.

\begin{table}[h]
\caption{Model configuration for NF-UNSW-NB15-v2 dataset}
\label{tab:params-nfunsw-v2}
\centering
\begin{tabular}{ll}
\toprule
\textbf{Parameter} & \textbf{Value} \\
\midrule
Scaler & MaxAbsScaler \\
Quantile threshold \( q \) & 0.95 \\
Anomaly rate parameter \( \nu \) & 0.1 \\
Learning rate \( \eta \) & 0.3 \\
\bottomrule
\end{tabular}
\end{table}

Compared to the original dataset, the configuration for NF-UNSW-NB15-v2 uses a slightly lower quantile threshold and higher values for both the anomaly proportion parameter and the learning rate. The threshold \( q = 0.95 \) indicates a more permissive decision boundary, likely needed to accommodate the increased variability and volume of anomalous flows in the extended dataset. The higher value of \( \nu = 0.1 \) suggests that the model expects a greater proportion of anomalies, aligning with the increase of anomalous flows present in this version. Additionally, the learning rate \( \eta = 0.3 \) enables faster adaptation, which can be advantageous in the presence of more dynamic or heterogeneous traffic patterns.

Finally, the comparative analysis between datasets suggests that the model generalizes well across different variants of UNSW-NB15. While the first version achieves slightly better accuracy, the extended version (v2) provides perfect anomaly coverage in terms of recall, likely due to the increased diversity and volume of anomalous flows.

\section{Conclusion}\label{sec:conclusion}

The exponential growth in the number of devices connected to the network, together with the sustained increase in cyberattacks in recent years, has highlighted the need for effective and adaptive security mechanisms. One of the main current challenges lies in detecting malicious traffic in networks, where the sheer volume of data generated by these devices makes it computationally unfeasible to analyze each packet individually. In this context, methods based on network flow analysis are particularly well suited.

In this work, we presented an anomaly detection model for network flows using an unsupervised online learning algorithm. This paradigm has proven to be especially effective in contexts where the dynamic nature of traffic and the scarcity of labeled data make the use of traditional supervised techniques impractical.

The main contributions include the development and implementation of a One-Class SVM---based model using the River library, as well as the integration of a preprocessing methodology tailored for continuous flow environments. The model was evaluated using two versions of the NF-UNSW-NB15 dataset, achieving remarkable performance metrics: accuracy above 98\%, a false positive rate below 3.1\%, and a recall of 100\% in the most advanced version of the dataset. This enables the detection of all anomalous flows while maintaining a very low false positive rate.

Moreover, the reduced processing time per flow---less than 0.033 ms---demonstrates that such solutions are feasible for deployment in real-time detection systems, even in infrastructures with limited computational resources. The model’s continuous adaptability allows it to maintain consistent performance in scenarios where legitimate traffic patterns evolve over time, making it capable of detecting zero-day anomalies.

It is worth noting that unsupervised models with online learning capabilities are still scarce compared to traditional batch learning approaches. This work makes a relevant contribution in this area, demonstrating both the technical feasibility and the competitive performance of such models in dynamic network scenarios.

Based on the results obtained, several lines of future research are opened. In particular, the incorporation of hybrid approaches that combine online and offline learning could further enhance the system’s detection capabilities \cite{odiathevar_online_2022, odiathevar_hybrid_2019}. 











\printcredits

\section*{Acknowledgements}

This research is a result of the CIBERLAB project (C083/23), carried out under the collaboration agreement between INCIBE and the University of León. This initiative is part of the Recovery, Transformation and Resilience Plan, funded by the European Union (Next Generation EU).

\bibliographystyle{cas-model2-names}

\bibliography{references}



\end{document}